\documentclass[11pt]{article}
\usepackage[margin=1.0in]{geometry}
\usepackage{graphicx}
\usepackage{epsfig}
\usepackage{amssymb,amsfonts}
\usepackage{mathrsfs}
\usepackage{epstopdf}
\usepackage{hyperref,color}
\usepackage{natbib}
\usepackage{setspace}
\usepackage{xypic}\usepackage[all,cmtip]{xy}

\usepackage{braket}
\usepackage{amsmath}

\newtheorem{defi}{Definition}

\begin{document}

\title{Spacetime is as spacetime does}
\author{Vincent Lam and Christian W\"uthrich\thanks{We wish to thank audiences in Golm (December 2014), Paris (March 2015), EPSA (September 2015), Geneva (October 2015, April 2017, and March 2018), Varna (May 2016), and Trieste (July 2017). We acknowledge support from the Swiss National Science Foundation (grant 105212\_169313) and from the John Templeton Foundation (grant 56314, performed under a collaborative agreement between the University of Illinois at Chicago and the University of Geneva). We are grateful to Lorenzo Cocco, Nick Huggett, Baptiste Le Bihan, Niels Linnemann, Keizo Matsubara, Tushar Menon, Daniele Oriti, James Read, David Yates, Alastair Wilson for discussions and comments on earlier drafts.}}
\date{27 April 2018}
\date{}
\maketitle

\begin{abstract}\noindent
Theories of quantum gravity generically presuppose or predict that the reality underlying relativistic spacetimes they are describing is significantly non-spatiotemporal. On pain of empirical incoherence, approaches to quantum gravity must establish how relativistic spacetime emerges from their non-spatiotemporal structures. We argue that in order to secure this emergence, it is sufficient to establish that only those features of relativistic spacetimes functionally relevant in producing empirical evidence must be recovered. In order to complete this task, an account must be given of how the more fundamental structures instantiate these functional roles. We illustrate the general idea in the context of causal set theory and loop quantum gravity, two prominent approaches to quantum gravity.
\end{abstract}


\section{Introduction}
\label{sec:intro}

The main research programs in quantum gravity (QG) tend to show that standard relativistic spacetime is not fundamental. The precise and different ways in which it is not fundamental depend on the particular quantum theory of gravity under consideration, but they all seem to suggest a radical picture according to which crucial spatiotemporal features (and possibly even space and time themselves) are not part of the fundamental physical ontology (\citealp{HuggettWuthrich2018}). 

This perspective raises two related families of philosophical worries (besides the many technical difficulties that the various QG approaches have to face). The first concerns the very possibility of empirical evidence, including the experimental confirmation of these theories themselves; after all, it seems that all experimental data ultimately always amount to some physical object, such as the `pointer' of some measurement apparatus, having a certain position in space at a certain time. Indeed, if space and time are necessary `preconditions' of theory confirmation in empirical science, then any theory denying the fundamental existence of spacetime undermines the very possibility of its own empirical justification. Consequently, such a theory would seem empirically incoherent (\citealp{HuggettWuthrich2013}). 

In most of the physics literature on quantum gravity, this challenge of empirical incoherence largely amounts to the usual constraint of consistency with the superseded theories: in particular, any theory of QG should recover in some appropriate regime the smooth relativistic spacetime picture of the theory of general relativity (GR).\footnote{Additionally, a theory of QG should of course come with an ontology that can accommodate evidence that goes beyond that of superseded theories, but such an ontology will in general not be of straightforwardly four-dimensional objects.} This consistency constraint is a central concern in all quantum gravity programs and may typically involve approximations and limiting procedures (\citealp{Wuthrich2017}). In this perspective, the issue is mainly a technical one. However, from a more philosophical point of view, the worry is that the consistency constraint is a necessary but not sufficient condition for the challenge of empirical incoherence to be met. To many, it remains unclear in what sense quantities such as space and time can emerge from a fundamental non-spatiotemporal ontology at all (\citealp{Maudlin2007a,LamandEsfeld2013}).

The second type of philosophical worries that may arise has to do with the very characterization of the possibly non-spatiotemporal physical ontology. First of all, in order to describe a plurality of physical entities in this context, a differentiation in spatiotemporal terms (according to which we have a diversity of fundamental physical entities in virtue of their distinct spacetime locations) is obviously not available within this framework. A related issue concerns spelling out compositional and mereological relations in non-spatiotemporal terms (cf.\ \citealp{Ney2015}).

Most importantly, it should be clarified what makes the non-spatiotemporal entities described by QG concrete physical entities, rather than merely abstract mathematical ones. The standard criterion for distinguishing the concrete from the abstract relies on spacetime itself: concrete entities are in spacetime, abstract ones are not. Clearly, such a spacetime criterion is just not available for characterizing a physical ontology of non-spatiotemporal entities. An alternative characterization of concrete entities involves some notion of causal efficacy: concrete physical entities as opposed to abstract mathematical ones can be considered as causally efficacious in some sense. However, at first sight, it seems far from obvious how to make explicit a precise notion of non-spatiotemporal causation (\citealp[\S 4.2]{LamandEsfeld2013}).

Thus, the two standard criteria for concreteness, which rely on spacetime and causation respectively, are not apt in the QG context (at least to the extent that this latter involves an ontology of non-spatiotemporal entities). However, they together suggest a mixed strategy that combines certain aspects of the two criteria: to focus on spacetime functions---that is, on spatiotemporal or `spacetime-like' roles, in some broad functional rather than narrowly causal sense---the fundamental entities may instantiate in favourable circumstances.

This paper aims to evaluate to what extent the tools of functionalism can help to alleviate the worries mentioned above, with particular attention to the threat of empirical incoherence. One central intuition we investigate is that spacetime need not be fully recovered in some strong ontological sense---to be explicated below in section \ref{sec:quantum ontology} and particularly in section \ref{sec:functionalism}---in order to provide the grounds for empirical evidence and everyday experience, but only certain functionally relevant features. We will argue that this functional approach offers a promising avenue towards meeting the conceptual challenge of the `emergence of spacetime' in QG.

Since the way in which spacetime is not fundamental in QG, and consequently the way in which it emerges at a higher level depends on the specific program under consideration, this functionalist strategy needs to be investigated in concrete cases. We focus in this paper on two important research programs in QG: causal set theory (CST) and loop quantum gravity (LQG). These two lines of research embody typical features that make the standard spacetime picture irrelevant at the QG level, mainly in that they offer a view of fundamental non-metrical discreteness in the first case, and quantum and combinatorial features in the second case. Although neither of these research programs is committed to spacetime functionalism of course, we will investigate to what extent a functionalist perspective would allow them to bridge the metaphysical gap between the structures they postulated and smooth classical spacetime as we find it in GR. 

Our functionalist project bears some similarities, as well as some dissimilarities, to related approaches in recent philosophy of physics. Perhaps most prominently, a functionalist strategy has been deployed in the context of non-relativistic quantum mechanics by David \citet[Ch.\ 6]{Albert2013} in defence of wave function realism and by David \citet{Wallace2012} in support of an Everettian interpretation. While their functionalism is concerned with recovering three-dimensional space and its manifest content from the wave function by means of a functional characterization of spatial dimensions, ours will be tasked with the emergence of four-dimensional spacetime, which renders their crucial use of dynamics inapt for the present project.\footnote{The functionalist approach about spacetime developed in this paper aims to constitute an interpretative strategy for understanding (the emergence of spacetime in) quantum gravity; in this sense it also differs from \citet{Chalmers2012}'s functionalism about spatiotemporal features centred on perception.}

Closer to our concern is the spacetime functionalism of Eleanor \citet{Knox2013,Knox2014,Knox2017}, which extends prior work by Harvey \citet{Brown2006}. Concerned with GR and rival classical theories, hers is a functionalism which concludes in a spacetime substantivalism; semantic quarrels aside, since we are interested in QG, where the fundamental structures are clearly less than fully spatiotemporal, we prefer to say that spacetime emerges from the fundamental level and that this emergence is precisely the relation that a functionalization is supposed to elucidate. Whether one ultimately wishes to be a realist or an eliminativist about spacetime is orthogonal to our concern here.

Let us illustrate this. For Knox, something `plays the spacetime role' and thus \emph{is} spacetime ``just in case it describes the structure of the inertial frames, and the coordinate systems associated with these'' (\citeyear[15]{Knox2014}; cf.\ also \citealp{Knox2013}). Since in GR it is the metric field which describes the structure of the inertial frames, the metric just is the spacetime (and hence the substantivalism). Since ``[t]here is nothing in the functional definition of spacetime to suggest that the realizer of the spacetime must be fundamental'' (ibid., 18), she claims that her functionalism can straightforwardly cope with the situation where spacetime emerges and GR is an effective, rather than a fundamental theory. Thus, the idea is that whatever the fundamental degrees of freedom, they give rise to that which realizes the role of inertial frames and hence spacetime; in other words, it assumes that the realizers are emergent themselves. This immediately raises the question what the relationship between these realizers and the fundamental degrees of freedom is supposed to be. Since this relationship is then not touched by Knox's spacetime functionalism, it is also not clarified by it.\footnote{Knox's functionalist is explicitly a \emph{realizer} form of spacetime functionalism, which identifies spacetime with the entity which instantiates the relevant functional role, rather than a \emph{role} spacetime functionalism, which identifies spacetime with the property of having certain properties that occupy the spacetime role. Although one may read our article as an expression of this latter view, we remain uncommitted between the two. In fact, we wish to leave open the possibility of resisting the dichotomy. We thank David Yates and an anonymous referee for pushing us on this point. For a discussion of this distinction in the context of spacetime functionalism, see \citet{Yates2018}, who offers a taxonomy of solutions to the problem of empirical incoherence, as well as an argument against the brand of realizer functionalism found in \citet{Chalmers2012}.} 

It is our declared goal to elucidate how spacetime can emerge from a non-spatiotemporal structure and what it might mean in a program in QG to secure this emergence. Thus, we wish to apply functionalism one level deeper, as it were: for us, the realizers of the relevant spacetime functions will be the fundamental degrees of freedom and their collective behaviour or structure. In fact, pace Knox, these realizers have to ultimately be placed at the fundamental level, even though the spacetime roles they play become apparent only at an effective scale. Furthermore, although we appreciate Knox's insistence on inertial frames, we want to enter our functionalist project with a more flexible understanding of what spacetime roles there are to be filled.

Before discussing to what degree certain spacetime roles can be functionally instantiated in CST (section \ref{sec:CST}) and LQG (section \ref{sec:LQG}), we will consider the (dis)analogies with the debate about the role of spacetime in the ontology of quantum mechanics (section \ref{sec:quantum ontology}) and with functionalist perspectives in the philosophy of mind and the philosophy of the special sciences (section \ref{sec:functionalism}). Conclusions follow in section \ref{sec:conc}.

\section{Quantum ontology and spacetime}
\label{sec:quantum ontology}

Some of the issues that quantum-gravitational theories `without spacetime' have to face (see section \ref{sec:intro}) already arise in the context of the debate on the ontology of the realist `interpretations' (or `versions') of quantum mechanics (QM). Main examples of such realist `interpretations' are Everettian or many-worlds theories (MW), Bohmian mechanics (BM) and dynamical (spontaneous) collapse theories, such as the theory of Ghirardi, Rimini and Weber (GRW). This section aims to see what can be learned from this debate for the QG context, highlighting the similarities and the differences between the cases of QM and of QG.

The just mentioned interpretations are realist with respect to QM in the standard sense that they take its central theoretical entity, namely the quantum state, or the wave function, to encode objective features of the physical world. A crucial part of the debate on the ontology of these realist interpretations of QM concerns the status of the wave function and the exact way in which it encodes objective features. 

At first sight, within this realist framework, the wave function is most naturally understood as a concrete physical object (namely, a physical field), very much like the electromagnetic field of Maxwell's theory. However, there is a huge difference between this latter field and the wave function of (many-particle) non-relativistic QM. Whereas the electromagnetic field lives on the usual 3-dimensional space (or 4-dimensional spacetime), the wave function is defined on the 3$N$-dimensional configuration space of QM, where $N$ is the number of particles under consideration---ultimately the total number of particles in the universe. From an ontological point of view, the term `configuration space' is a misnomer here, since there actually are no configurations of particles in standard 3-dimensional space in this picture of the world but only the wave function living on this very high-dimensional space and evolving in time. According to this straightforward realist conception of the wave function (called `wave function realism' or `wave function monism'; see \citealp{Albert1996,Albert2015,NeyandAlbert2013}), standard 3-dimensional space, which constitutes the natural framework for classical physics and for empirical evidence in general, is simply not part of the fundamental physical ontology.\footnote{It is of course possible to include standard 3-dimensional space in the fundamental ontology of QM \emph{in addition} to the configuration space of the wave function; but this move would considerably inflate the ontology and raise the issues of the relationship between these two spaces and of how things in these different spaces can seemingly interact.} The challenge of how to account for empirical evidence in this context is thus issued, since empirical data are intuitively understood as ultimately always involving (what looks like) objects located in 3-dimensional space (such as the pointer of a measurement apparatus) and evolving in time. 

In this sense, there is a similarity between the case of QG without spacetime and the wave function realist understanding of QM: in both cases, the fundamental ontology of the theory does not explicitly include ordinary 3-dimensional space and entities localized in ordinary 3-dimensional space, thereby posing a possible threat for the empirical coherence of the theory.\footnote{Elaborating on previous considerations by John S. Bell, \cite{Maudlin2007a, Maudlin2010} explicitly raises the threat of empirical incoherence as an argument against wave function realism.} Within the framework of wave function realism, the worry is that the wave function living on a very high-dimensional configuration space is not the right sort of entity to ground the 3-dimensional experimental evidence that is commonly understood to empirically confirm QM in the first place. The fundamental ontology of wave function realism does not contain any entities localized in ordinary 3-dimensional space---what Bell calls `local beables', i.e., the entities that exist according to the theory and that ``can be assigned to some bounded spacetime region'' (\citeyear[53]{Bell1987})---and so cannot rely on such localized entities or local beables to account for experimental evidence. 

In the context of wave function realism, an obvious strategy to address this issue of empirical incoherence is to deny that local beables are required in the fundamental ontology in order to ground experimental evidence. \citet{Albert1996, Albert2013} precisely deploys this strategy, relying on a functionalist understanding of ordinary objects localized in 3-dimensional space: these latter are understood in terms of their causal and functional roles, that is, in terms of the causal and functional relations they enter into.\footnote{``And the thing to keep in mind is that what it is to be a table or a chair or a building or a person is---at the end of the day---\emph{to occupy a certain location in the causal map of the world}. The thing to keep in mind is that the production of geometrical appearances is---at the end of the day---a matter of \emph{dynamics}.'' \cite[127, emphases in original]{Albert2013}.\label{QuoteAlbert2013}} Albert then crucially argues that the dynamics of the wave function is such that it can play the functional role of 3-dimensional situations (which are encoded in the form of the Hamiltonian)---instantiating the right sort of functional relations---that constitute the basis for empirical evidence (see also \citealp[ch.6]{Albert2015}).    

As mentioned in the introduction, the functionalist strategy that this paper aims to investigate in QG is partly inspired by the use of functionalist tools in the context of wave function realism.\footnote{See also \citet[ch.2]{Wallace2012}'s functional emergence of macroscopic structures (the `many worlds') in his Everett interpretation of QM.} However, beyond the similarities between wave function realism and the QG theories `without spacetime', it is important to underline the fundamental differences between the two cases, in order to better grasp the specific nature of the problem in QG. Most importantly, the QM context allows for spacetime-based ontologies that constitute alternatives to wave function realism in a way that is not straightforwardly available or legitimate in the context of QG. 

Indeed, building on the work of Bell, research in the last decades has made clear that each of the three main realist versions of QM (i.e.\ BM, GRW, MW) can be understood in terms of an ontology of local beables (with a non-local dynamics), that is, an ontology of material entities localized in (bounded regions of) ordinary 3-dimensional space (or 4-dimensional spacetime), such as particles, point-like events (`flashes') or a matter density field.\footnote{It has been argued in \citealp{Allorietal2011} that this is the case even for some versions of MW.} Such an ontological framework of local beables has been called `primitive ontology' in the recent literature (see \citealp{Allorietal2008}): it is primitive in the sense that it is what makes up everything else (including empirical evidence)---everything is ultimately \emph{constituted} by elements in the primitive ontology. It is also primitive in the sense that it is postulated from the start, rather than read off from the mathematical structure of the theory (as, to some extent, wave function realism is). 

The arguments in favour of the primitive ontology approach to QM mainly involve its explanatory structure: it aims at providing an explanatory framework within which familiar macroscopic objects localized in 3-dimensional space, such as those involved in measurement outcomes, can be understood in terms of (more) fundamental entities (particles, flashes, matter density field) that are also localized in 3-dimensional space. In this sense, ordinary 3-dimensional space (or 4-dimensional spacetime) provides an explicit and natural connection between the primitive ontology and what can be observed at the familiar macroscopic level---such an intuitive connection is simply lacking within wave function realism, which has to explain ordinary macroscopic objects that are (or seem to be) localized in 3-dimensional space in terms of a fundamental entity that is not. The appeal to functionalism aims precisely at providing such an explanatory connection in the case of wave function realism. Establishing such a connection is pressing for the wave function realist, as the alternative strategy of positing an ontology of local beables---\emph{to the extent that such an ontology is available at all}---can be argued to be more intuitive and natural, as an inference to the best explanation. In particular, it is the `constitution' or `building blocks' aspect of such an ontological framework (everything physical being ultimately made up of fundamental local beables) that can be seen as more intuitive and natural than the functionalist perspective. Although the devil may be in the details,\footnote{The explanatory structure of the primitive ontology approach is not as simple and straightforward as sometimes suggested: indeed, the local beables alone cannot ground (`constitute') ordinary macroscopic objects and their behaviour without taking into account the wave function, which therefore remains a key element in the explanatory structure (see e.g. \citealp[\S 5]{NeyandPhilipps2013}).} the intuitive force of this argument for an ontology of local beables should not be underestimated.\footnote{For a recent collection of papers on the debate between the primitive ontology approach and wave function realism in QM, see \citet{NeyandAlbert2013}.}          

Such an ontological framework of local beables fundamentally relies on spacetime and on localization with respect to spacetime. This obvious fact highlights a crucial difference about the role of spacetime in the QM and QG contexts. In the former case, spacetime is fixed and non-dynamical (though possibly curved) and thus not part of the dynamical physical systems described by the theory, whereas in the latter case (the geometry of) spacetime itself is considered as a dynamical system, and therefore not fixed a priori. The various research programs in QG actually investigate the very nature and status of spacetime itself, as well as its relationship to matter, with the possibility of spacetime being non-fundamental considered very seriously (see section \ref{sec:intro}). In this context, it seems neither legitimate nor methodologically adequate to merely postulate from the start an ontological framework according to which the physical systems described by QG are localized with respect to some fundamental spacetime structure or in fundamental spatiotemporal terms (at least, in the standard sense of `spacetime'; see \citealp[\S 7 and \S 9]{Lam2015}).\footnote{To the extent that the notion of `local beables' requires an unambiguous notion of localization, a tension may already arise in the context of classical GR, where there is no external, fixed background with respect to which physical systems can be unambiguously localized.} Similarly, the intuitive appeal of the `constitution' or  `building blocks' perspective seems to vanish in a non-spatiotemporal context (what it can mean for something spatiotemporal to be constituted by something non-spatiotemporal is obviously far from intuitive). 

This leads us to another aspect of the difference between the QM and QG cases: there simply is no issue about time in the ontology of QM. Both the wave function and the local beables evolve in time, i.e.\ take time for granted within the ontological framework of QM. The dynamics of the wave function actually plays a central role in Albert's functionalist account of the everyday picture of the world within wave function realism: the wave function on configuration space evolves in time in such a way that it can play the functional role of 3-dimensional situations (see Albert's quote in footnote \ref{QuoteAlbert2013}). On the contrary, there is no such standard temporal evolution of fundamental physical systems available in the QG context where space \emph{and} time generally do not possess any fundamental status. The difference between QM and QG on the issue of time is well illustrated by \citet[\S 7]{Ney2015}, who---while defending wave function realism against the charge of empirical incoherence---considers that QG theories without time do face a ``legitimate threat of empirical incoherence'', since standard confirmation theories take confirmation as a fundamentally ``diachronic process'': time, and above all, change are necessary conditions for empirical confirmation. The idea here is that time is crucial for the empirical coherence of a theory in a way that space is not. The functionalist strategy that we investigate in this paper rejects this distinction between space and time with respect to empirical coherence: it aims at applying the same functionalist understanding to both the spatial and temporal features that are relevant for empirical confirmation and coherence, as well as for the account of the standard spacetime picture of the world.\footnote{The functionalist strategy investigated in this paper is the kind of ``reconstruction project'' considered by \citet{Healey2002} as a way for the contemporary, physics-based Parmenidean to avoid empirical incoherence and self-refutation.} We now turn to the conceptual framework of functionalism that will allow us to do that. 

\section{Functionalism in philosophy of science and functionalism about spacetime}
\label{sec:functionalism}

Functionalism in philosophy actually designates a variety of different linguistic and metaphysical conceptions mainly in the philosophy of mind and the philosophy of the special sciences, which all emphasize an understanding of relevant predicates and properties in terms of the role they play in the context where they are defined. For instance, within such a functionalist framework, a mental property is determined by its causal role; in fact, it is \emph{identified} with its causal role within the network of mental activities. In general, however, functional properties need not be exclusively characterized in causal terms, even if they typically are so characterized in the cases of mental properties and special sciences properties; in the context of QG, we need to consider a broader notion of functional role, one that does not rely on spatiotemporal notions in any way. 

The aim is to investigate to what extent certain generic features of functionalism (in its metaphysical garb) can help us to better grasp the relationship between the more fundamental level of QG and the smooth spacetime level of GR (and of most of physics). Indeed, in both the philosophy of mind and the philosophy of the special sciences, the functionalist emphasis on the notion of functional role alongside the usual notion of composition provides a fruitful understanding of the links between (allegedly) qualitatively different levels, such as the mental (or biological, chemical, etc.) level and the physical level.\footnote{The functionalist and constitution/composition perspectives can be complementary in certain cases (as arguably in the context of the special sciences), but stand in opposition in others (as in the context of the emergence of spacetime).} Analogously, the issue of the emergence of relativistic spacetime can be considered from a functionalist perspective on the relationship between (what seems to be) qualitatively different levels, that is, the non-spatiotemporal level of QG and the spatiotemporal level of GR.    

As a general framework, let us consider the model of functional reduction of higher-level properties to lower-level ones, as developed in the context of the philosophy of mind and the philosophy of the special sciences (see for instance \citealp[101f]{Kim2005}).\footnote{At first sight, it might seem a bit awkward to consider a reductionist framework in order to better understand the \emph{emergence} of spacetime. Two things on this: first, the aim is to focus on the role of functionalism in the functional reduction model and second, emergence (at least in a certain sense) and reduction need not stand in opposition, see \citet{Butterfield2011a, Butterfield2011b} and \citet[\S2]{cro16}. In fact, philosophers should take note that when physicists speak of the ``emergence of spacetime'' in QG, they do not mean to state that spacetime cannot be `reduced' to the underlying non-spatiotemporal physics they are proposing.\label{ftn:reductionism}} Such a functional reduction typically involves two mains steps:\footnote{\citet[102]{Kim2005} considers three steps; we simply take the third to be part of the second.}
\begin{description}
\item[(FR-1)] The higher-level properties or entities, which are the target of the reduction, are `functionalized', that is, they are given a functional definition in terms of their causal or functional role. 
\item[(FR-2)] An explanation is provided of how the lower-level properties or entities can fill this functional role. 
\end{description}
This is very schematic, but the details of particular functional reduction models in the philosophy of mind and the philosophy of the special sciences are not needed for our present purposes. In particular, at this stage at least, we need not engage with the consequences of functional reduction for the general philosophical debate on reductionism. The aim here is first to investigate to what extent this functionalist scheme can provide a fruitful conceptual framework for understanding the relationship between smooth spacetime and possible non-spatiotemporal QG entities---most importantly, a conceptual framework where the specific challenges raised by this relationship (such as empirical incoherence) can be met. Whether and to what extent this relationship should be qualified as an instance of reduction or emergence in the philosophical sense is a secondary issue in the QG context; for now, we just highlight the important fact that (as already noted in footnote \ref{ftn:reductionism} and following \citealp{Butterfield2011a, Butterfield2011b}) the two need not stand in opposition, so that the relationship encoded by {\bf(FR-1)} and {\bf(FR-2)} above may in some circumstance well be one of functional \emph{emergence} in that the higher level entities exhibit some novel and robust behaviour relative to the entities at the more fundamental level.

In the case of the relationship between smooth spacetime and the non-spatiotemporal entities of QG, a crucial element of this general framework of functional reduction (or functional emergence) is the `functionalization' of spacetime and spatiotemporal properties: in the first step {\bf(FR-1)}, spacetime and spatiotemporal properties need to be understood in terms of their functional roles, so that it can then be shown how the QG entities can fill these functional roles in the second step {\bf(FR-2)}. As a consequence---and this is a novel perspective brought by this functionalist framework---the focus not only lies on the QG entities that might give rise to classical spacetime, but also on classical spacetime itself, which needs to be functionally re-interpreted. More precisely, this `functionalization' is primarily aimed at the spacetime properties that are relevant for empirical confirmation, such as spacetime localization. Quite generally, the functional role of spacetime localization---e.g. the functional role of the property of `being located in spacetime region $\mathcal{O}$'---can be understood in terms of its role in the relevant theoretical structure. We will discuss the specific functional role of spacetime localization in the dynamical context of GR in more detail in the next section. Before that, in the remainder of this section, we would like to discuss some general, but important issues about functional emergence as encoded in {\bf(FR-1)}--{\bf(FR2)}.

The functional understanding of higher-level properties suggested by {\bf(FR-1)} allows different (kinds of) lower-level properties to play the same (kind of) functional role, i.e. different (kinds of) `realizers' for the same (kind of) higher-level property. This is called `multiple realizability' in the philosophy of mind and the philosophy of the special sciences, where it constitutes a central topic---in particular the consequences of multiple realizability for reductionist theories. As already mentioned above, we are not directly concerned with the philosophical debate on reductionism here. Whether or not it is compatible with a genuine reduction,\footnote{Again, we tend to side with Butterfield in opposing the general argument from multiple realizability against reduction (\citeyear[\S 4.1.1]{Butterfield2011a}).} it is seen in general as a virtue of functionalism that it allows---indeed, accounts for---multiple realizability.

The liberal attitude of functionalism with respect to what fills the functional roles may raise the worry that functionalism provides too weak a metaphysical understanding of higher-level properties in the sense that functional roles may not exhaust the nature of these properties. This general worry is widespread in the philosophy of mind, where it can take various forms. For example, a forceful objection considers that functionalist theories do not have the resources to capture the qualitative and phenomenal aspects (often called `qualia') of certain mental properties (or mental states), e.g. those linked with conscious experience. Without entering into the details of this hotly debated topic in the philosophy of mind, the point for our purpose is clear: the worry is that the functionalization of spacetime properties, as suggested in {\bf(FR-1)}, may leave out crucial (\emph{spatiotemporal}) aspects of these properties, rendering the corresponding functional emergence unconvincing and unhelpful, or at least incomplete (as a consequence, such an account would not bring anything new beyond the mere technical consistency constraint and would not be able to meet the philosophical challenges that are linked to the emergence of spacetime, see section \ref{sec:intro}). 

In order to make the charge more vivid, we can perhaps imagine `spacetime zombies'. Suppose there are two worlds exactly alike in their fundamental substance, the `stuff' they are made out of, yet they differ in the properties which are exemplified by that stuff in these two cases. In the first world, just as in ours, the fundamental substance has both non-spatiotemporal as well as spatiotemporal properties. In this world, then, the fundamental substance's nature directly includes spatiotemporal qualities and it is therefore no question that we have spacetime. In the second world, which is an exact duplicate of the first world in all non-spatiotemporal respects, the fundamental substance does not exemplify spatiotemporal properties. Although the fundamental physics and all the functionally characterizable spatiotemporal appearances will be identical in the two worlds, the latter world lacks a decisive spatiotemporal quality and thus merely contains a `spacetime zombie'. Since such a spacetime zombie world seems numerically distinct and, it is insisted, metaphysically possible, there must also be directly spatiotemporal properties exemplified in a world which encompasses spacetime. Thus, in particular, there cannot be a functional reduction of spacetime to something less than fully spatiotemporal. 

It is not clear, however, how much traction the `qualia' concern really gets in the spacetime case as compared to the philosophy of mind. As \citet{Knox2014} puts it, ``[w]here the fan of qualia has introspection, the fan of the [spacetime] container has only metaphor'' (16). The nature and status of the evidence in favour of qualia may be equivocal, but the alleged ineliminable intrinsically spatiotemporal but ineffeable quality of a spacetime substance remains positively elusive. What could remain of that quality once we have accounted for all relevant spatiotemporal features such as (relative) localization as captured by the relative spatial and temporal order, their metric valuations in spatial distances and temporal durations, and perhaps more? Moreover, one may reasonably resist the idea that such spacetime zombie worlds are metaphysically possible.

A specific version of this worry about functionalism has been raised in the context of wave function realism (section \ref{sec:quantum ontology}). \citet{Ney2015} argues that, although functionalist accounts ``make some headway in indicating how the wave function may possess some of the features of three-dimensional objects'', they are ``not adequate to demonstrating how the wave function may actually constitute these types of things'' (3116).\footnote{\citet{Ney2015} actually defends wave function realism, in particular against the threat of empirical incoherence, by rejecting the need of local beables for empirical confirmation in the first place. Since her strategy only applies to space (but not to time, see above section \ref{sec:quantum ontology}), it is not suited to the QG context.} She then considers holograms as an example. The hologram of an object seems to possess many features of the object it is an hologram of, but that does not entail that ``it actually constitutes'' the object in question. Thus, she concludes, what is needed is not just a compelling account ``of how a particular evolving wave function may have features in common with three-dimensional objects..., but a compelling story as to how a wave function is capable of actually \emph{constituting} such objects'' (ibid., our emphasis). Ney's constitution objection (called the ``macro-object problem'') straightforwardly transfers to the QG context: functionalism does not seem to possess the resources to account for how something non-spatiotemporal can constitute things with spacetime features, such as being localized in space and time, partly because the functional roles of the things with spacetime features may not exhaust their (spatiotemporal) nature.

Two interrelated considerations may block this constitution objection in the context of QG. First, the notion of constitution in the strong (spatiotemporal) sense involved in the objection may not be needed (and relevant) in this context. Indeed, the framework of QG may well use mereological tools without relying on spacetime, and may well appeal to general, non-spatiotemporal compositional relations (together with some non-spatiotemporal differentiation of the basic entities of QG that functionally---and non-spatiotemporally---`compose' the things with spacetime features). Second, functionalism about (standard, smooth) spacetime is committed to the view that there is nothing about (standard, smooth) spacetime---more precisely: nothing about relevant smooth spacetime properties---beyond their functional role. Spacetime just is what plays the spacetime role, faithful to the slogan `spacetime is as spacetime does'. As a consequence, anything playing the right functional role will realize the corresponding spacetime feature or the corresponding spatiotemporal `situation'. Even if such a functionalist view of spacetime---of relevant spatiotemporal features---might sound surprising or unintuitive at first, it must be made clear that there is nothing \emph{incoherent} or plainly wrong about it, especially given its strong epistemic anchorage: indeed, we mainly have epistemic access to physical entities through what they do, which is precisely what is encoded in their functional role.\footnote{In this perspective, all science is mainly about functional roles; this is clearly expressed by \citet[233]{Dennett2001}: ``Functionalism is the idea enshrined in the old proverb: handsome is as handsome does. Matter matters only because of what matter can do. Functionalism in this broadest sense is so ubiquitous in science that it is tantamount to a reigning presumption of all science.'' All science includes spacetime physics.} Note that in Ney's example above, the hologram of an object does not fill the right functional role of the object in all its aspects (e.g.\ it does not interact with other objects in the right way), hence it does not functionally realize---and so a fortiori not `constitute'---the object.    

The epistemic relevance of functional roles, including spatiotemporal ones, is central to the functionalist reply to the challenge of empirical incoherence. The functionalist perspective, and in particular functional emergence, allows one to preserve---and, to some extent, to ground---the intuitive epistemic primacy of certain spatiotemporal features or spatiotemporal `situations', such as a (macro-)object having a certain position in space at a certain time (e.g.\ a `pointer' in a measurement context). As a consequence, the argument that empirical confirmation requires certain spacetime features (such as spacetime localization, local beables) constitutes no threat to theories `without spacetime': as long as the basic entities in QG play the right functional roles (in the right circumstances), such as that of localization, empirical confirmation just proceeds the usual way in the context of QG, and empirical coherence is not under threat because spacetime is not part of the fundamental ontology of the theory. To the extent that this latter is empirically successful (this is crucial!), the functionalist perspective guarantees that there is no specific reason to be skeptical or instrumentalist about the theory just because it does not postulate some standard spacetime structure in its foundations---in other words, the traditional epistemological debate can be held in the usual way.\footnote{See \citet{LamOriti2018} for a more detailed discussion; thanks to Nick Huggett for valuable exchanges on this point.}  

So, from a functionalist point of view, nothing remains beyond showing how the fundamental degrees of freedom can collectively behave such that they appear spatiotemporal at macroscopic scales \emph{in all relevant and empirically testable ways}. This turn out to be a hard task in quantum gravity. Functionalism can be seen as the assertion that once this task is completed, no unfinished business lingers on. Using a terminology from philosophy of mind, functionalism would then amount to the denial that there is a `hard problem' beyond the `easy problem' of the emergence of spacetime.\footnote{This distinction was imported into the context of the emergence of spacetime by \citet{LeBihan2018}.}

Crucial to this functionalist strategy is of course to functionally define the relevant spacetime features {\bf(FR-1)}, and to show concretely how these functional roles can be filled {\bf(FR-2)}. To the extent that there is no complete theory of quantum gravity yet, this second step can only be sketched. The next two sections aim to articulate the main lines of this functionalist strategy in the concrete cases of two research programs in QG, namely causal set theory (section \ref{sec:CST}) and loop quantum gravity (section \ref{sec:LQG}).

\section{Functional emergence of spacetime in causal set theory}
\label{sec:CST}

Causal set theory (CST) is founded upon the principle that the fundamental structure underlying relativistic spacetime is discrete, consists of a partially ordered set of otherwise featureless elementary events, where the partial order is induced by a relation of causal precedence. It is motivated by a theorem in GR which was proven in its strongest form by David \citet{Malament1977} and states that given the causal structure of a causally sufficiently well behaved general-relativistic spacetime, the dimension, topology, differential structure and metric (up to a conformal factor) of the spacetime is determined. In other words, in much of GR the causal structure determines the geometry of a spacetime, but not its `size'. It is under the impression of this remarkable theorem that the developers of CST have made the causal structure the backbone of their approach to QG. More precisely, the theory contends that the fundamental structure is a causal set $\mathcal{C}$, i.e., an ordered pair $\langle C, \prec\rangle$ of a set $C$ of events and a relation $\prec$, which partially orders $C$.\footnote{A partial order is a binary relation defined on a domain, which is reflexive, antisymmetric, and transitive.} Furthermore, it is assumed that causal sets are discrete structures, i.e., that for all $a,c\in C$, the cardinality of $\{b\in C| a \prec b \prec c\}$ is finite.\footnote{For a review on CST, see \citet{Sorkin2005,Dowker2013}; see \citet{Wuthrich2011} for a philosophical assessment.} It should be noted at the outset that while the ambition of the program is to produce a \emph{quantum} theory of gravity, the theory remains classical to date.

In CST, the fundamental relation is causal, rather than temporal; temporal relations are supposed to emerge from the structure of the causal set.\footnote{More precisely, it is a primitive technical relation of causal connectability quite different from the more usual philosophical understanding of causation.} So strictly speaking, there is no time in CST. More impressionistically, $\prec$ looks like a discrete, (special-)relativistic time without metric relations such as durations. Space, on the other hand, does not exist in CST. Any subset of events in any causal set with any claim to represent a spatial slice at a time would have to be such that its elements are pairwise unrelated by $\prec$---for otherwise, one of the elements would precede the other---and thus by construction be utterly structureless: no topological, affine, conformal, let alone metric structure, and no natural dimensionality. So how does relativistic spacetime emerge from these structures? Can we precisely state necessary and sufficient conditions for causal sets to give rise to relativistic spacetimes?

A fully satisfactory answer to this question must cover various aspects. First, it should clarify what the basic set-up is: are we trying to understand the relationship between theories, or rather between their models, or some subsets of the set of their models? Also, it is important to deliver not just conditions necessary for a mathematical derivation, but also sufficient for the demonstration of the physical salience of the emergence.

So, how should we conceive of the basic situation? The traditional answer in CST proceeds along the following general recipe. First, we are seeking a mapping from causal sets to relativistic spacetimes, which formally captures the relation between the former and the latter. More precisely, we are looking for an injective map $\varphi: C \rightarrow M$, where $M$ is the manifold of a relativistic spacetime $\langle M, g_{ab}\rangle$ with metric $g_{ab}$. The mapping should be such that it becomes evident how the salient spatiotemporal features of the codomain---the spacetime---arise from the properties of the domain---the fundamental causal set. It thus relates single models of the two theories and indicates the direction of their ontological dependence or ground from fundamental to emergent. By mapping fundamental events into events of the emerging spacetime, it identifies the basal events with some of the higher-level events. Due to the discreteness of the causal set, almost all of the higher-level events will not be dignified by being selected as images under $\varphi$ of one of the fundamental events, and hence the map will be non-surjective. The general idea then is to show how the salient spatiotemporal or geometrical properties of the specific emergent spacetime at stake are functions of the properties of the specific causal set. In a further step, it has to be shown that this procedure can generically be applied to the two sets of models from the two theories, i.e., for most physically reasonable spacetimes we can find a pairing map and articulate how its relevant features arise from the properties of the underlying causal set. But much more needs to be said about how that is supposed to work.

Generally, we would expect such a mapping from causal sets to spacetimes to be many-to-one: each higher-level spatiotemporal property can generally be implemented in many different fundamental ways, just as in philosophy of mind one would expect mental properties to be multiply realizable by the grounding physical properties (section \ref{sec:functionalism}). Furthermore, we should expect some corrections to GR; after all, it would be hard to justify replacing GR with a more fundamental theory if there was no divergence at all between the two theories. For instance, given the antisymmetry (and the transitivity) of $\prec$, it may well be that CST only reproduces the sector of GR of spacetimes without closed timelike curves. Also, models in GR with high energy or matter density such as they appear in the early universe in the standard model of classical relativistic cosmology may be ruled out by `quantum effects'. In general, one may hope that those sectors of GR containing `physically unreasonable' spacetime models (such as, e.g., when energy conditions are violated) would not (and should not) be recoverable from CST.

CST as we have formulated it so far is extremely permissive in that it admits way too many models which have no prayer of giving raise to a physically reasonable spacetime---or just to a spacetime for that matter. So some additional conditions need to be imposed on the general set-up to restrict the class of admissible models adequately. First, causal sets have to be `sufficiently large' to carry enough structure to mimic even just a small piece of continuum spacetime. As \citet{BombelliEal1987} suggest, the causal set has to consist of at least $10^{130}$ elements in order to give rise to a spacetime sufficiently vast to describe our world. However, it turns out that almost all `sufficiently large' causal sets permitted by the kinematic axiom are so-called `KR orders', i.e., degenerate partial orders of only three `layers' or `generations' of elements such that all pairs of elements of subsequent generations stand in $\prec$. It is clear that such a causal set cannot successfully describe a realistic spacetime and cannot generically realize the right sort of spacetime functions. The way causal set theorists have sought to avoid the `KR hordes' is by imposing additional conditions in the form of `dynamical' rules. Which kind of dynamical rules are most promising remains under investigation.\footnote{Cf.\ \citet{ridsor99} for the most widely discussed approach.} It is at this point that advocates of CST hope to turn their theory into a {\em quantum} theory: these dynamical rules will ultimately have to be appropriately quantum in nature.

\begin{figure}
\centering
\epsfig{figure=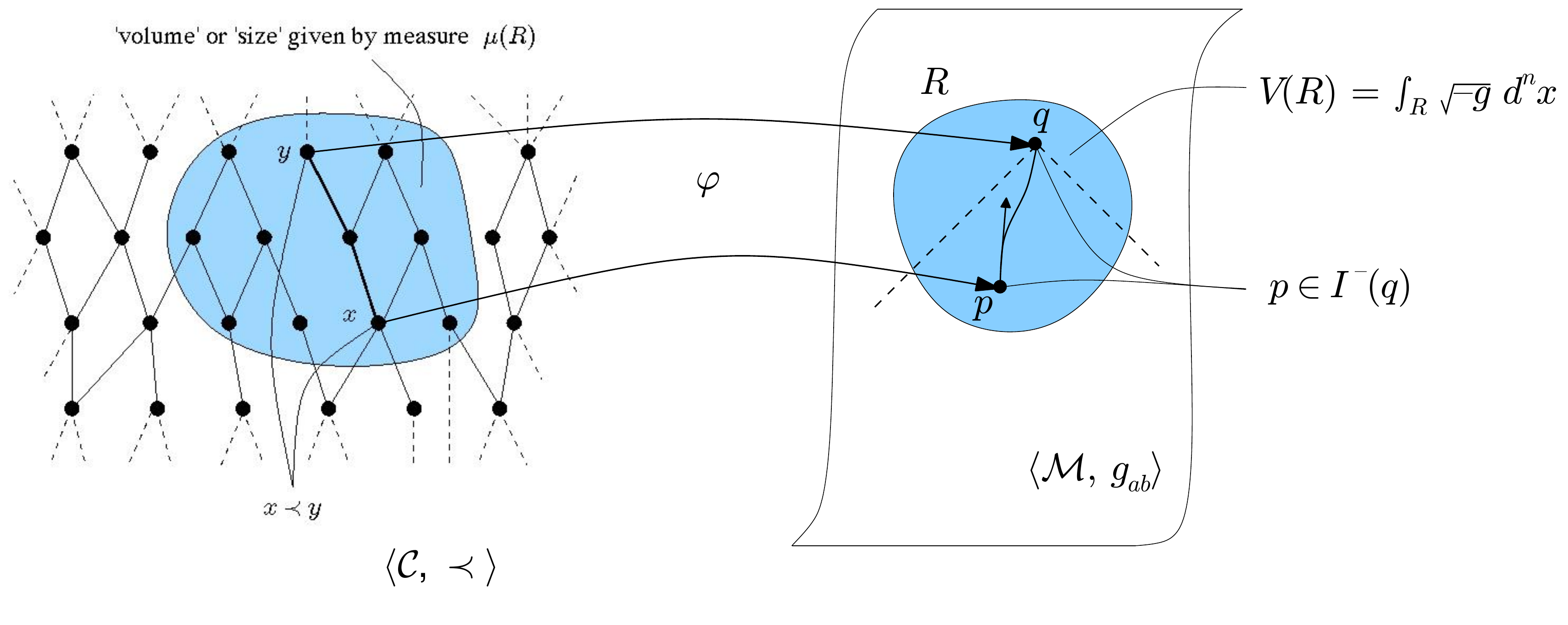,width=\linewidth}
\caption{\label{fig:faithful} Finding a map $\varphi: C \rightarrow M$ such that the image looks like a manifold.}
\end{figure}

Unfortunately, even if the KR catastrophe is avoided, there is no guarantee that the dynamically admissible causal sets will be `manifoldlike', i.e., give rise to manifolds of reasonably low dimensionality or that is has Lorentz signature or a non-pathological causal structure, and that it can thus perform the spacetime functions, such as the localization function, required of it. A causal set thus has to play a `manifold functional role', i.e., it has to play some of the roles the manifold plays for the smooth spacetime. In other words, it has to be relevantly like a manifold. Thus, we will call a causal set $\langle C, \prec\rangle$ {\em manifoldlike} just in case it is `well approximated' by a relativistic spacetime $\langle M, g_{ab}\rangle$. In order to get any traction, the expression in quotes needs a more rigorous formulation. The standard way of articulating this idea, which goes back to \citet{BombelliEal1987}, uses the notion of `faithful embedding' of a causal set into a spacetime.\footnote{There are at least two other suggestions for how to address this task. First, as proposed to us in conversation by David Meyer, one could attempt to define a measure `$P(C|M)$' to express the probability that a given causal set $\langle C, \prec\rangle$ arises from random sprinkling into a manifold $M$, and then use Bayes's Rule to obtain a probability distribution $P(M|C)$ for Lorentzian manifolds. This suggestion has not been worked out. Second, as proposed in unpublished work by Luca Bombelli and Johan Noldus, one could proceed by understanding both causal sets and (future and past distinguishing) Lorentzian manifolds as `causal measure spaces' and using a Gromov-Hausdorff function $d_{GH}(\cdot, \cdot)$ (or the Gromov-Wasserstein function if we have a probability measure) to give a measure of the closeness or similarity between such spaces.} Along these lines, the definition can be precisified as follows---cf.\ also Figure \ref{fig:faithful}:
\begin{defi}\label{def:manifoldlike}
A causal set $\langle C, \prec\rangle$ is said to be {\em manifoldlike} just in case there exists a `faithful embedding', i.e., an injective map $\varphi : C\rightarrow M$, where $M$ is the manifold of a spacetime $\langle M, g_{ab}\rangle$, such that
\begin{enumerate}
\item the causal relations are preserved, i.e.\ $\forall a, b\in C, a \prec b$ if and only if $\varphi(a) \in J^-(\varphi(b))$;
\item $\varphi (C)$ is a `uniformly distributed' set of points in $M$;
\item $\langle M, g_{ab}\rangle$ does not have `structure' at scales below the mean point spacing.
\end{enumerate}
\end{defi}
It should be fairly clear that this is a definition in the functionalist spirit we are invoking: not the full manifold with its metric structure has to be recovered---the third clause purges the excess structure of the smooth manifold---, but a sufficiently fine-grained network of events, which correctly represents the causal structure of the underlying causal set. It is this network that then has to be shown to be sufficient to perform the spacetime functions required of it. 

Suppose now that the causal set at stake is manifoldlike and world enough for our universe. Two further conditions are necessary to complete the setting. First, causal sets need to be approximated not just by any spacetime, but by `physically reasonable' relativistic spacetimes. In other words, the spacetimes entering Definition \ref{def:manifoldlike} are sufficiently well-behaved relativistic spacetimes. Second, if a causal set is approximated by a spacetime, then the approximating spacetime is `approximately unique'. This is what causal set theorists call the ``Hauptvermutung'' of their programme, and it is imposed because if one and the same causal set could give rise to distinct spacetimes on different occasions, we would lose any meaningful way of saying that it emerges from an underlying causal set. In other words, this condition should ensure that one and the same causal set cannot play inconsistent spacetime roles. In terms of the above characterization of `manifoldlike', this means, more precisely, that if two relativistic spacetimes $\langle M, g_{ab}\rangle$ and $\langle M', g_{ab}'\rangle$ both approximate a given causal set in that there exist faithful embeddings $\varphi : C \rightarrow M$ and $\varphi' : C \rightarrow M'$, then they are `approximately isometric'. To demand exact isometry would be too strict, since a causal set with its ultimately discrete structure should not be held to capture minute metric differences between the two spacetimes if they are below the scales that could register in the causal set and thus should be functionally inert. A continuum spacetime contains `too much information', as it were, for a causal set to be able to fully accommodate that, and more than is needed to execute all relevant spacetime functions. As the isometry thus does not have to be exact, we should, of course, define the expression `approximately isometric'. We will not attempt this here, but submit that the causal set programme will have to deliver an appropriate characterization, together with a proof of the ``Hauptvermutung''. 

Would the completion of this task fully explain how relativistic spacetimes emerge from causal sets? No: the only properties of the causal set and the spacetime that the setting has related to one another so far are their causal structures, which have to be essentially the same. Malament's theorem is a theorem of GR, but not of CST. Although it provided essential motivation, we cannot just assume that the causal structure and the discreteness at the fundamental level give us a full-fledged emerging spacetime. We insist, however, that it is not necessary that the fundamental causal set possesses, in a narrow and literal sense, spatial or temporal properties. Instead, we maintain that in order to establish the emergence of relativistic spacetimes from causal sets, it is sufficient to complete the two tasks outlined in section \ref{sec:functionalism}. In the terms used there, it first must be determined which are the `spatiotemporally salient' aspects of a high-level structure that need to be accounted for, and hence functionalized {\bf(FR-1)}. These aspects are geometric and include spatiotemporal localization, spatial distance, temporal duration, topology, and similar aspects of the geometry of spacetime. At the second step of the functional reduction, {\bf(FR-2)}, one needs to show how the fundamental entities can fill these geometric roles. The mapping sought after in the preceding paragraph forms part of this second step. Importantly, what any such `derivation' needs to get right are only the functionally relevant features of relativistic spacetimes. Although there is no guarantee that all these features can be recovered, the preservation of the causal structure is imposed upon those embeddings precisely in order to secure this. 

So how can we extract geometric information from the fundamental causal sets such that this information can then be used to relate them to the smooth spacetime of GR, thus exemplifying the functional reduction we are arguing for? In order to complete this task in practice, some concrete means of extracting such information must be constructed. Much work has been done in this direction, and it continues to be an active research area. For instance, researchers have attempted to define the spatial topologies and spacelike distances in terms native to the fundamental causal set, showing that at least in some contexts these topologies and distances closely correspond to the relevant aspects of relativistic spacetimes in which these causal sets can be embedded.\footnote{\citet{Major2006,Major2007} construct a spatial topology by `thickening' the `spatial' antichains of a causal set, i.e., subsets of events which are pairwise incomparable, and exploiting the arising causal structure; \citet{Rideout2009a} offer the most penetrating analysis for the case of spacelike distances.} 

As an illustration of functional reduction in CST, let us consider dimension. Dimension, arguably and in response to step {\bf(FR-1)}, is functionally relevant for instance through its many appearances in emergent dynamical patterns which are encoded in the laws of non-fundamental theories, such as Newton's law of universal gravitation. How can we identify some property or combination of properties of the underlying causal set that would be indicative of the dimensionality of the `smallest' manifold into which it can be faithfully embedded? The standard definition of the `dimension' of a partial order such as a causal set defines it as the minimum integer $n$ such that the partially ordered set can be embedded in $\mathbb{R}^n$ with the so-called `coordinate order', i.e.\ with the ordering relation $\leq$ which obtains between points in $\mathbb{R}^n$ just in case their standard coordinates all satisfy the usual `less or equal' relation \citep{Brightwell1997}. This recipe yields a natural number $n$ as a candidate for the dimension of the emerging spacetime. However, although perhaps intuitive, this notion of dimension is in general hard to compute, only gives a very weak upper bound and, most importantly, is not straightforwardly transferable to the case of embeddings into relativistic spacetimes. In order to overcome that deficiency, one can introduce the `Minkowski dimension, i.e., the dimension of the lowest-dimensional Minkowski space into which a causal set can be embedded \citep{Meyer1988}. Analytically unresolvable, mathematicians have developed promising numerical methods to estimate the Minkowski dimension of causal sets based on properties of the fundamental structure alone. If thoroughly successful, this would complete step {\bf(FR-2)}. 

The question of how a fundamental causal set can give rise to the dimension of an emerging space thus finds a functional resolution. Similar stories can be told regarding some other geometric properties of spacetimes,\footnote{For a more detailed discussion also of the cases of spatial distance and topology, cf.\ \citet[Ch.\ 3]{HuggettWuthrich2018}.} though we certainly do not pretend that a functional reduction of relativistic spacetimes to causal sets is almost accomplished or imminent. Much work remains to be done. But what we do want to insist on is that it is precisely this kind of work that needs to be completed, {\em and nothing else on top}. Once all features of spacetimes as identified in step {\bf(FR-1)}, such as dimension in the example above, have been given a functional understanding in a second step {\bf(FR-2)}, the emergence of relativistic spacetimes from an underlying quantum-gravitational structure has been achieved. Let us turn to our second illustration.

\section{Functional emergence of spacetime in loop quantum gravity}
\label{sec:LQG}

At the heart of the research program in loop quantum gravity (LQG) lies the dynamical nature of spacetime as encoded in GR: the received view (at least from the perspective of LQG) is that the spacetime metric and the gravitational field are aspects of the same dynamical structure, and, as a consequence, there is no external, fixed, non-dynamical background with respect to which physical entities can be dynamically considered (the diffeomorphism invariance of the theory is taken to reflect this dynamical nature).\footnote{Many physicists at the origin of the LQG program have emphasized the importance of this dynamical understanding of spacetime (and the related notion of `background independence') at the classical GR level, in particular for the foundations of LQG---Carlo Rovelli perhaps more prominently than anyone else (see his \citeyear{Rovelli2004} textbook on LQG). This received view can be qualified in different ways, which need not concern us here.} To some extent, the dynamical nature of spacetime can be considered as the root of the `disappearance of spacetime' in QG (at least in the QG approaches, such as LQG, that take this dynamical nature of spacetime seriously): to the spacetime (metric) structure of the classical theory corresponds a quantum structure at the quantum (gravitational) level---since it is a dynamical entity---whose features make any (standard) spacetime interpretation difficult.

Now, the interesting thing is that the dynamical nature of spacetime explicitly suggests a functional understanding of certain crucial spacetime features already at the classical (GR) level---in particular, spacetime localization---thereby motivating the functional emergence framework. Indeed, a radical consequence of spacetime itself being dynamical is that spacetime localization at the classical level amounts to localization with respect to a dynamical entity that has no privileged dynamical status in general. This seems to imply that localization, which plays such a central role for the issue of empirical coherence (see section \ref{sec:intro}),\footnote{That it plays such a central role is the reason why we mainly focus on the functional emergence of spacetime localization here; however, ultimately, one would also need to show how other spacetime features relevant for the issue of empirical coherence functionally emerge in the appropriate way.} may primarily be understood as relational and dynamical rather than fundamentally spatiotemporal: localization in this GR sense is to be understood in terms of the (cor)relations among the dynamical entities of the theory rather than with respect to some fixed, non-dynamical spacetime background.\footnote{\citet{Rovelli1999} highlights the conceptual novelty of this dynamical notion of localization within GR.} This suggests a functional understanding of localization in the sense of certain functional relationships being satisfied.  

The definition of physical (gauge-, diffeomorphism-invariant) `coordinates' in GR nicely illustrates this functional understanding of localization. Physical localization can be defined with respect to four independent scalar functionals $\phi^K[g]$ ($K = 1, \ldots, 4$) of the gravitational field. These scalar functionals play the role of physical coordinates with respect to which physical entities can be localized (they play the `localizing role' of spacetime points).\footnote{These physical coordinates (sometimes called `pseudo-coordinates' to distinguish them from mere mathematical coordinates) have been originally defined in the context of generic (i.e.\ without non-trivial symmetries) pure gravitational (i.e.\ vacuum) solutions of the Einstein field equations (e.g.\ see \citealp{Komar1958}), but the idea could in principle be extended to generic solutions with matter.} For instance, in this dynamical (and diffeomorphism-invariant) perspective, a physical field $F$ is not genuinely localized with respect to some fixed spacetime background: strictly speaking, $F(x)$, $x \in M$, where $M$ is a spacetime manifold, does not represent any physical event, since it is not diffeomorphism- (or gauge-) invariant, and so does not express any genuine spacetime localization. The physical field $F$ is rather localized with respect to other dynamical fields: $F(\phi^K)$, which is diffeomorphism- (or gauge-) invariant, primarily encodes a dynamical and functional notion of localization, in the sense of certain functional relationships between dynamical entities. As such, these functional relationships are not inherently spatiotemporal.\footnote{Although this dynamical conception of localization seems to naturally suggest a relationalist ontology at the classical GR level, it does not rule out substantivalism, since the scalar fields $\phi^K$ are functionals of the metric and can therefore be understood as encoding the spacetime structure (i.e., they may receive a spacetime interpretation). However, this debate is secondary for the aim of this paper: whatever is the preferred spacetime ontology at the classical level, localization in the GR context can be functionally understood, and as such this very functional understanding does not involve spacetime notions (in gauge-theoretic terms, it amounts to a `gauge fixing', see \citealp[fn 22]{Earman2002}).} In the schematic terms of the functional emergence framework proposed in section \ref{sec:functionalism}, these relationships can actually be understood as \emph{functionally defining} the (`higher-level', GR) property of `being localized in a certain spacetime region', thereby providing an instance of how the first step {\bf(FR-1)} can be implemented. 

To the extent that localization can be implemented in various ways within GR---with respect to different dynamical entities, none being privileged in principle---the corresponding functional definition can take various forms.\footnote{For instance, the dynamical `localizing role' can be instantiated in principle by different specific (non-degenerate) matter distributions within GR (such as matter scalar fields).} The `localizing function' in these different instantiations involves at its core the notions of coincidence and contiguity: the functional role of localizing a physical entity (what it means for a physical entity to be localized) crucially involves coincidence and contiguity relations.\footnote{This has been clearly underlined by \cite{Einstein1916} and his discussion of ``space-time coincidences''; more recently, see Rovelli's emphasis on the notion of contiguity for the GR (dynamical) understanding of localization (e.g. \citealp{Rovelli1999}). Furthermore, \citet[\S2.4.3]{RovelliandVidotto2015} discuss an interesting analogy between contiguity and interaction, suggesting the idea, to which we will return below, that QG interactions can give rise to---\emph{can play the role of}---spatiotemporal contiguity in appropriate circumstances.} As a consequence, in schematic terms, the second step {\bf(FR-2)} crucially involves how LQG entities (or LQG properties) can instantiate coincidence and contiguity relations in an appropriate context.    

At the kinematical level, the fundamental LQG entities are described by spin network states, which form a basis of the relevant kinematical Hilbert space of LQG.\footnote{Since, for the time being, LQG is better  developed and understood at the kinematical level, part of the discussion is focused on this level. Dynamical aspects are considered later in this section.} They correspond to quantum states of (3-)space (regions) and of the (3-)gravitational field. Spin network states are represented by combinatorial graphs carrying irreducible group ($SU(2)$) representations on their links and nodes. A crucial aspect of the physical relevance of the spin network states comes from the fact that they are eigenstates of the area and volume operators (built from the classical geometrical expressions), which turn out to have discrete spectra. This result is naturally interpreted as encoding some fundamental discreteness: the spectral analysis of these operators suggests that the nodes of the spin networks represent `quanta of volume' (or `atoms of space')\footnote{There are various reasons for being very cautious with this evocative analogy about `atoms of space' or `atoms of spacetime', see below and section \ref{sec:conc}.} and the links represent `quanta of area'. Although radically different from the continuous spacetime structure of GR, this fundamental discreteness does not, as such, preclude any spatiotemporal interpretation at the LQG level, although some properties of relativistic spacetimes such as their smoothness are definitely lost. In particular, the links can be naturally understood as instantiating contiguity relations (also called `adjacency relations') between the `atoms of space'.\footnote{See \citet[189]{Rovelli2004}: ``Two chunks of space are contiguous if the corresponding nodes are connected by a link $l$. In this case, there is an elementary surface separating them [\ldots] The graph $\Gamma$ of the spin network determines the adjacency relation between chunks of space.'' However, \citet[\S D]{Rovelli2011} warns that the geometrical (and spatiotemporal) picture of the discrete structures associated with spin networks ``should not be taken too literally'', since there are different, not necessarily ontologically equivalent, geometrical understandings of spin networks. Cf.\ also \citet[\S2.3]{HuggettWuthrich2013}, \citet[\S 2.2]{LamandEsfeld2013} and \citet[\S2]{Wuthrich2017} for a discussion on the ways in which spin networks are less than fully spatiotemporal.} 

This spacetime understanding actually runs into difficulties because of two further important features of LQG. First, generic LQG states are quantum superpositions of spin network states, so that generic states are not associated with a single discrete structure, but with a superposition thereof, making any straightforward spacetime interpretation difficult. Second, the fundamental adjacency relations represented by the spin network links need not---and in general do not---correspond to spatiotemporal, metrical contiguity in the standard GR sense: two connected nodes at the LQG level may correspond to events in GR spacetime that are arbitrarily far away (in the GR metric sense) from one another.\footnote{This feature is called `disordered locality', see \citet{MarkopoulouSmolin2007}, as well as \citet[\S 28.5.3]{Smolin2009} and references therein; for a philosophical discussion, see again \citet[\S 2.3]{HuggettWuthrich2013} and \citet[\S 2.1]{Wuthrich2017}.\label{ftn:disorderedlocality}} In this perspective, the fundamental spin network connectivity cannot be directly interpreted in standard spatiotemporal terms; most importantly, it does not give rise at the LQG level to any notions of localization and locality that come close to the standard ones---the ones that are so central in the threat of empirical incoherence and in the argument for \emph{local} beables. In functionalist terms, the standard `localizing function' is not instantiated at the LQG level in general. Now, the crucial point is to show that spin networks \emph{can} play a localizing role in the standard sense, in particular at the GR level: the higher-level property of `being localized in a certain spacetime region' (in the standard, GR sense) will then be considered to \emph{functionally emerge} from the LQG properties of spin networks.

From a technical point of view, understanding how a standard notion of spatiotemporal locality may arise from the fundamental LQG description is part of the general issue of the classical limit of the theory, which still remains one of the key open problems of LQG. Moreover, to the extent that it is a quantum theory, a full understanding of the classical limit of LQG ultimately requires addressing the quantum measurement problem. Our aim here is neither to address the technical difficulties linked to the classical limit of LQG nor to take a stance on the measurement problem.\footnote{To some extent, the measurement problem needs to be addressed for a quantum theory to have a clear ontology; in particular, the precise ontological meaning of quantum superpositions will depend on one's stance with respect to the measurement problem.} Rather, the aim is to show how a functionalist perspective can alleviate the specific conceptual difficulties related to the emergence of the standard notions of space and time from a fundamental level where they may be absent. 

In order to discuss a concrete example, we apply this functionalist perspective to the so-called `weave state approach' to the classical limit.\footnote{As we will briefly mention below, the general conceptual framework of functionalism applies to other approaches to the classical limit as well, e.g. in terms of coherent states; it may also be relevant in the case where spacetime features may arise in the continuum or thermodynamic (rather than classical) limit (\citealp{Oriti2014}).} Weave states are specific (`semi-classical') spin network states that have a large-scale classical (spatiotemporal) behaviour in a precise sense: they are eigenstates of the geometrical (volume, area) operators with eigenvalues approximating the corresponding classical values as determined by the classical metric. In more technical terms, a spin network state $\ket{S}$ is a weave state for a classical (3-)metric $g$, if we have, at a scale $l\gg$ Planck length $l_P$ and for a sufficiently large 2-surface $\mathcal{S}$ and a sufficiently large 3-region $\mathcal{R}$, and up to small corrections of the order $\mathcal{O}(l_P/l)$ \cite[\S 6.7.1]{Rovelli2004}:
\begin{equation}\label{eq:weave}
\begin{aligned}
\hat{\mathbf{A}}(\mathcal{S}) \ket{S} = (\mathbf{A}[g, \mathcal{S}] + \mathcal{O}(l_P/l)) \ket{S},\\
\hat{\mathbf{V}}(\mathcal{R}) \ket{S} = (\mathbf{V}[g, \mathcal{R}] + \mathcal{O}(l_P/l)) \ket{S}
\end{aligned}
\end{equation}
where $\hat{\mathbf{A}}(\mathcal{S})$ and $\hat{\mathbf{V}}(\mathcal{R})$ are the area and volume operators corresponding to $\mathcal{S}$ and $\mathcal{R}$, and $\mathbf{A} [g, \mathcal{S}] = \int |d^2\mathcal{S}|$ and $\mathbf{V} [g, \mathcal{R}] = \int |d^3\mathcal{R}|$ are the classical area and volume as determined by the classical metric $g$. Now, in a functionalist perspective, the spin network state $\ket{S}$ instantiates the metrical function---plays the metrical role---of the (classical) volume $\mathbf{V}$ and area $\mathbf{A}$ in the appropriate regime (at a large scale and for a large region $\mathcal{R}$ and a large surface $\mathcal{S}$) in the sense that it determines the same volume $\mathbf{V}$ and area $\mathbf{A}$ as the classical metric $g$.\footnote{Cf.\ \citet[\S4.2]{Wuthrich2017} for a more detailed account.}

This functionalist understanding naturally accounts for the fact that (\ref{eq:weave}) does not determine $\ket{S}$ uniquely: since (\ref{eq:weave}) only puts a constraint on properties averaged over a large region and a large surface (compared to the length scale $l$), many different spin network states can play the same `averaged' metrical role. From a functionalist point of view, this seems at first sight akin to a case of multiple realizability (see section \ref{sec:functionalism}), where different lower-level properties---here, represented by the spin network states---can play the same functional role, that is, can instantiate the same higher-level properties---here, classical volume and area. However, the standard philosophical notion of multiple realizability involves multiple realization by different types (kinds) of physical states or properties or configurations; and here it is not clear to what extent we really have different types of LQG configurations---different types of spin network states---differently realizing the same type of metrical function.\footnote{We are grateful to Nick Huggett for highlighting this point to us; see \citet{LamOriti2018} for a more detailed discussion.}  

The important point for our discussion is that this functionalist perspective provides a clear (namely, functional) sense in which metrical (and more generally, spatiotemporal) properties can emerge in principle from a non-metrical (non-spatiotemporal) level: the (non-metrical) spin network states of the lower level of LQG can play certain metrical functional roles at the higher level of GR (in the appropriate regime), e.g.\ as encoded in (\ref{eq:weave}).\footnote{The metrical meaning encoded in (\ref{eq:weave}) actually depends on the interpretation of quantum operators (observables) in general, that is, ultimately on the solution to the measurement problem that is privileged; e.g. Bohmians and Everettians (whatever their conceptions precisely amounts to in the QG context) do not attribute the same meaning to (\ref{eq:weave}). We believe that the broad functionalist framework suggested here can be further specified in the context of the different ontological stances one can have with respect to the quantum formalism and the measurement problem: this is because the functionalist perspective on the emergence of spacetime does neither rely on, nor exhaust, the \emph{exact} ontological nature of the lower level of LQG.\label{ftn:measurement proble}}

Classical relativistic spacetime (or the classical metric) has of course many different functional roles, but not all these roles may be instantiated by the lower-level properties of LQG, or at least not directly. This is crucially illustrated by the issue of disordered locality (see footnote \ref{ftn:disorderedlocality}) at the semi-classical level. Indeed, the conditions (\ref{eq:weave}) defining weave states do not preclude any mismatch between spin network connectivity at the LQG level and standard spacetime locality at the GR level \cite[]{MarkopoulouSmolin2007}. If this mismatch is sufficiently drastic for some weave state/spacetime pair, then this weave state would not instantiate any standard `localizing function', while still playing the metrical roles of volume and area as encoded in (\ref{eq:weave}). Most importantly, standard (`spatiotemporal') locality would then not be instantiated and the threat of empirical incoherence would arise, since the theory would not be able to account for the local beables that ground our experimental evidence. From a functionalist perspective, the threat of empirical incoherence would arise in such a case not because of some in principle conceptual problems linked to the emergence of spacetime from a non-spatiotemporal level, but rather because the right functional role---namely, the localizing function---is not instantiated. Indeed, it is expected that disordered locality is rather mild for (single) weave states, so that these latter actually do play the right localizing role at the higher level of GR, thereby avoiding the threat of empirical incoherence. Again, the functionalist framework naturally accounts for the fact that different weave states with various configurations of (mild) disordered locality can play the same localizing role: novel (ultimately local!) empirical predictions may actually arise from this (mild) disordered locality (such as predictions of the fluctuations in the cosmic microwave background, among others; see \citealt[\S 28.5.3 and \S28.5.4]{Smolin2009}).\footnote{As already mentioned above, the low-energy limit of LQG is still far from being fully understood; indeed, many technical difficulties remain. For instance, single weave states are very specific semi-classical states; since quantum superposition is a generic feature of any quantum theory, superpositions of semi-classical states are actually to be expected, in particular with consequences for (disordered) locality. It has been suggested that decoherence effects may play a role here (with certain LQG degrees of freedom considered as `environmental', see the philosophical discussion in \citealt[\S 4.2]{Wuthrich2017}), but concrete implementations and control over quantum corrections are still wanting. These are technical and empirical issues that need to be further investigated. Functionalism obviously does not aim at addressing them, but rather aim at providing a clear conceptual framework for the emergence claims.\label{ftn:low-energy limit}}    

So far we have only considered the kinematical part of LQG, which is better developed and understood; according to the standard understanding of the theory, this means that we have only been concerned with the (functional) emergence of \emph{space} rather than spacetime. We conclude this section with a few remarks about the dynamical part of the theory and the (functional) emergence of \emph{spacetime}. 

In its canonical setting, the dynamics of LQG is encoded in the so-called Hamiltonian constraint (the LQG counterpart of the infamous Wheeler-DeWitt equation), whose definition is extremely tricky and for which no complete solution has been found so far (these difficulties partly find their roots in the fact that there is no external time parameter with respect to which the dynamics unfolds---hence the claims about the absence of time at the LQG level). Dynamical generalizations of spin networks in a covariant setting---the so-called `spin foams'---are increasingly being studied as an alternative approach to the dynamics of LQG. Indeed, spin foams can be intuitively understood as higher-dimensional (Feynman-like) graphs (`2-complexes') describing the `evolution' of spin networks (`1-complexes'). The dynamics is then understood in terms of transition amplitudes---more precisely: sums over spin foam amplitudes---between spin networks (or for single `boundary' spin networks). 

Spin foams resist any standard spacetime interpretation for reasons very similar to the ones invoked for spin networks.\footnote{There is actually a deep analogy between (single) spin foams and Feynman graphs, where any (discrete) spacetime picture attached to (single) spin foams have much the same `virtual' status as the geometrical picture of Feynman graphs (away from the classical limit; see the philosophical discussion in \citealt[\S 2.2]{LamandEsfeld2013}). This Feynman analogy is exploited in the framework of group field theory (GFT), which is a quantum field theory (QFT) over a group manifold---in some sense, a kind of QFT of spin networks (see \citealt{Oriti2014}, \citealt{LamOriti2018}).} The functionalist perspective discussed above in the context of the classical limit of spin network states (in terms of weave states) applies equally well to the classical limit of spin foam amplitudes (in terms of coherent states, see \citealt[Ch.\ 8]{RovelliandVidotto2015}):\footnote{As in the spin network case, the classical limit scheme here still faces many issues---see footnotes \ref{ftn:measurement proble} and \ref{ftn:low-energy limit}.} in the classical limit, spin foams play the spacetime role of a Regge (simplicial) discretization, that is, play the role---in particular, the localizing role---of a spacetime lattice, very much like in lattice field theory. Of course, the crucial difference with standard lattice field theory is that dynamical entities (spin foams), which do not themselves explicitly possess standard spatiotemporal properties, play the role---in particular, the localizing role---of the spacetime lattice in the appropriate regime (localization is then dynamically and functionally instantiated in the sense that the right functional relationships among dynamical entities are instantiated).\footnote{The Regge (simplicial) discretization then converges to classical relativistic spacetimes in the continuum limit. Note that it is important to distinguish the two kinds of limits (classical and continuum) here; in particular, it might be the case that the two do not commute and that interesting (spatiotemporal?) properties emerge in the continuum or thermodynamic limit of LQG or GFT (again, see \citealt{Oriti2014}; see also \citealt[13.2]{RovelliandVidotto2015}).} 

Similarly to the spin network case, the strength of this functionalist perspective on the emergence of standard spacetime comes from the flexibility with respect to what exactly instantiates the relevant spacetime roles. There are two aspects to this flexibility. First, it allows for relevant spacetime roles to be instantiated by different spin foams, which may differ with respect to other spacetime aspects (functions), in particular leaving room for new predictions (see the issue of disordered locality above). Second, the spacetime functional roles may not exhaust the nature of the entities (e.g.\ spin networks, spin foams) that instantiate these roles, leaving room for further metaphysical work based on the future developments of the theory. Indeed, if there is nothing about (classical, standard) spacetime (at the GR level) beyond the relevant spacetime functional roles (`spacetime is as spacetime does'), it might well be the case that the fundamental QG structures possess other features (besides being spacetime realizers) as captured by the (future) QG theories; in order to be considered as empirically meaningful---as \emph{physical} as opposed to `merely' mathematical---these (non-spatio-temporal) features must have some bearing on (some of) the relevant spacetime functional roles that are instantiated by the QG structures in the right circumstances (for example, non-geometrical phases being related to geometrical ones through some phase transition, e.g.\ see \citealt{Oriti2014}, \citealt{LamOriti2018}).

\section{Conclusions}
\label{sec:conc}

Functionalism in the context of the emergence of spacetime in quantum gravity can be characterized as a {\em rejection} of the demand that more is needed---beyond empirical success---than showing how the fundamental structure, whatever it may turn out to be, can fulfill the role or the function of spacetime, by showing how it can give rise to all the relevant properties of space and time. We believe that to date, no research program in quantum gravity has shown this. However, in order to progress on that task, it is also important to recognize just what it is that needs to be established. The central claim of this essay asserts that to require that the proposed reductions somehow show how some ill-defined qualitative features of spacetime are obtained from the fundamental non-spatiotemporal ontology, or even insist that such a reduction of spacetime `qualia' is in principle impossible, is misguided. Instead, it suffices to identify those aspects of spacetime which are necessary to support the physics we do and the world as we experience it, and then show how these arise, at least to a well-controlled approximation, from the fundamental physics of quantum gravity.

From a metaphysical point of view, the functionalist perspective articulated in this article can be understood as a move away from the `constitution' or `building blocks' heuristics according to which, in the case of quantum gravity, spacetime and spatiotemporal entities are constituted by, or made up from, quantum gravitational building blocks. The primitive ontology (or local beables) approach to the ontology of quantum mechanics is a paradigmatic example of this heuristics at work in a context other than quantum gravity. However valuable for physical and metaphysical theorizing in other contexts, this heuristics acts, when taken too seriously in the present context, as the source of conceptual worries concerning the constitution of spacetime and spatiotemporal entities from non-spatiotemporal, and hence qualitatively improper, elements.\footnote{We believe this heuristics is also partly behind the physicists' somewhat loose talk about the `atoms of spacetime' in quantum gravity.} In particular, one might think, based on this constitution heuristics, that theories of quantum gravity face an irresolvable problem of empirical incoherence which could only be addressed by admitting appropriately localized entities into one's fundamental ontology. We have argued that this heuristics is misguided, as the relationship between spatiotemporal and quantum gravitational structures is best understood in terms of some relevant \emph{functional} (rather than \emph{constitutive}) roles these latter instantiate in appropriate circumstances. Spacetime is as spacetime does. No more, no less.


\bibliographystyle{apa}
\bibliography{Ref}

\end{document}